\documentclass[preprint2]{aastex}
\begin{document}

\title{Subaru Deep Spectroscopy of the Very Extended Emission-Line
Region of NGC 4388: Ram Pressure Stripped Gas Ionized by the Nuclear
Radiation
\footnote{Based on data collected at the Subaru Telescope, which is operated by the
          National Astronomical Observatory of Japan.}}

\author{
Michitoshi Yoshida\altaffilmark{2,3},
Youichi Ohyama\altaffilmark{4},
Masanori Iye\altaffilmark{2},
Kentaro Aoki\altaffilmark{4},
Nobunari Kashikawa\altaffilmark{2},
Toshiyuki Sasaki\altaffilmark{4},
Kazuhiro Shimasaku\altaffilmark{5},
Masafumi Yagi\altaffilmark{2},
Sadanori Okamura\altaffilmark{5},
Mamoru Doi\altaffilmark{6},
Hisanori Furusawa\altaffilmark{4,5},
Masaru Hamabe\altaffilmark{7},
Masahiko Kimura\altaffilmark{8},
Yutaka Komiyama\altaffilmark{4},
Masayuki Miyazaki\altaffilmark{5},
Satoshi Miyazaki\altaffilmark{4},
Fumiaki Nakata\altaffilmark{5},
Masami Ouchi\altaffilmark{5},
Maki Sekiguchi\altaffilmark{9}
AND
Naoki Yasuda\altaffilmark{10},
}

\vspace {3cm}

\altaffiltext{2}{Optical and Infrared Astronomy Division, National Astronomical Observatory,
Mitaka, Tokyo 181-8588, Japan.}

\altaffiltext{3}{Okayama Astrophysical Observatory, National Astronomical Observatory,
Kamogata, Okayama 719-0232, Japan; yoshida@oao.nao.ac.jp.}

\altaffiltext{4}{Subaru Telescope, National Astronomical Observatory of Japan, 650 North
A'Ohoku Place, Hilo, HI 96720, USA.}

\altaffiltext{5}{Department of Astronomy, University of Tokyo,
Tokyo 113-0033, Japan.}

\altaffiltext{6}{Institute of Astronomy, University of Tokyo, Mitaka,
Tokyo 181-8588, Japan.}

\altaffiltext{7}{Department of Mathematical and Physical Sciences,
Japan Women's University, Bunkyo-ku, Tokyo 112-8681, Japan.}

\altaffiltext{8}{Department of Astronomy, Kyoto University, Kyoto 606-8502, Japan.}

\altaffiltext{9}{Institute for Cosmic Ray Research, University of Tokyo, Kashiwa, Chiba 277-8582, Japan.}

\altaffiltext{10}{Astronomical Data Analysis Center, National Astronomical Observatory, Mitaka, Tokyo 181-8588, Japan.}


\begin{abstract}

We report here the results of deep optical spectroscopy of the very
extended emission-line region (VEELR) found serendipitously around the
Seyfert 2 galaxy NGC~4388 in the Virgo cluster using the Subaru
Telescope. The H$\alpha$ recession velocities of most of the filaments
of the region observed are highly blue-shifted with respect to the
systemic velocity of the galaxy. The velocity field is complicated,
and from the kinematic and morphological points of view, there seem
to be several streams of filaments: low velocity filaments, with
radial velocities $v$ with respect to the systemic velocity of
NGC~4388 $\sim -100$ km s$^{-1}$, high velocity ($v \sim -300$ km
s$^{-1}$) filaments, and a very high velocity ($v \sim -500$ km
s$^{-1}$) cloud complex. The emission-line ratios of the VEELR
filaments are well explained by power-law photoionization models with
solar abundances, suggesting that the Seyfert nucleus of NGC 4388 is
the dominant ionization source of the VEELR and that the VEELR gas has
moderate metallicity. In addition to photoionization, shock heating
probably contributes to the ionization of the gas. In particular, the
filaments outside the ionization cone of the Seyfert nucleus are
mainly excited by shocks. The predicted shock velocity is $\sim 200$
-- 300 km s$^{-1}$, which is comparable to the velocities of the
filaments. We conclude that the VEELR was formerly the disk gas of NGC
4388, which has been stripped by ram pressure due to the interaction
between the hot intra-cluster medium (ICM) and the galaxy. The
velocity field and the morphology of the VEELR closely resemble
snapshots from some numerical simulations of this process. In the case
of NGC 4388, the ram pressure-stripped gas, which is normally seen as
extended \ion{H}{1} filaments, happens to be exposed and ionized by the
radiation from the AGN, and so can be seen as optical emission-line
gas.

\end{abstract}


\keywords{
galaxies: clusters: general ---  galaxies: evolution --- 
galaxies: individual (NGC 4388) --- galaxies: kinematics and dynamics --- 
galaxies: Seyfert --- 
intergalactic medium
}


\section{INTRODUCTION}

Determination of the environmental effects on the evolution of galaxies has been
one of the main subjects of extragalactic astronomy. It is widely
believed that phenomena such as morphology segregation and color
evolution of galaxies in clusters of galaxies provide important clues
to the nature of these environmental effects.

Morphology segregation of galaxies in clusters has been well known
since the 1970s (Oemler 1974; Melnick and Sargent 1977), --- early-type galaxies are the dominant population at the central region of a
cluster, while late-type galaxies are preferentially distributed at
the outskirts. Dressler (1980) found a good correlation between galaxy
morphology and the number density of galaxies in the local universe and
generalized the morphology segregation of cluster galaxies in terms of
a density -- morphology relationship. Postman and Geller
(1984) found that this relationship extends over six orders of
magnitude in galaxy density.

Meanwhile, Butcher and Oemler (1978, 1984) reported observational
evidence of the increase of blue galaxy fraction at high redshift
clusters, which is known as the ``Butcher-Oemler effect'' (Butcher and
Oemler 1978, 1984). In addition, recent deep imaging studies of high redshift
clusters performed with the {\it Hubble Space Telescope} have revealed
that the majority of the blue galaxy population of high-$z$ clusters
consists of star-forming normal late-type disk galaxies (e.g., Dressler
et al. 1994, 1997; Couch et al. 1998; Fasano et al. 2000). This
population is thought to be made up of field galaxies captured by
clusters, in the context of the hierarchical structure formation
scenario within a cold dark matter (CDM) cosmology, which is the most
successful model of the universe. These findings suggest that galaxies
captured by a cluster should change their morphology and color as
they fall into the cluster center (Abraham et al. 1996; Oemler,
Dressler \& Butcher 1997; Poggianti et al. 1999).

Several mechanisms for driving the morphology and color evolution of
galaxies in clusters of galaxies have been proposed. These include
successive fast, shallow encounters of galaxies (``galaxy
harassment'') (Moore et al. 1996; Moore et al. 1998); close galaxy --
galaxy interactions, such as mergers (Walker, Mihos, \& Hernquist 1996;
Kauffmann \& Charlot 1998; Okamoto \& Nagashima 2001); removal of the
halo gas by interactions with the hot intra-cluster medium (ICM)
(Larson, Tinsley \& Caldwell 1980; Bekki, Couch \& Shioya 2002); or
ram pressure stripping of the disk gas by the hot ICM (Abadi et al.
1999; Quilis, Moore \& Bower 2000; Vollmer, Cayatte, Balkowski, \&
Duschl 2001 (hereafter referred to as VCBD); Schulz and Struck 2001
(hereafter referred to as SS01)). It is still not clear which of these mechanisms plays
the greatest role in the evolution of cluster galaxies, although there have been many
observational and theoretical studies. Nevertheless, it
is clear that the rapid and drastic consumption of the interstellar matter
of a galaxy as it passes though a cluster is the key to its evolution.
Thus, detailed observational studies of the sites at which
violent gas consumption is occurring, such as large-scale outflows or strong
starbursts, would help us to understand this problem.

Recently, Yoshida et al. (2002; hereafter referred to as YOS02) found
a very large region of ionized gas extending around a Seyfert 2 galaxy
in the Virgo cluster, NGC 4388. This very extended emission-line
region (the ``VEELR'') has a size of $\approx 35$ kpc and is located
preferentially toward the northeastern side of the galaxy. The region
consists of many filaments or clouds, with a typical size of $\sim
100$ pc. The total ionized gas mass of the VEELR is $\sim 10^5$
M$_\odot$. The [\ion{O}{3}]$\lambda$5007/H$\alpha$ emission-line
intensity ratios of the filaments, and the observation that the ratios decrease
monotonically with distance from the galactic center,
suggest that the primary ionization source of the region may be the
Seyfert nucleus of NGC 4388. YOS02 proposed two possible hypotheses
concerning the origin of the VEELR gas: it may be either (1) the tidal
debris of a past minor merger or (2) the ram pressure-stripped gas due
to the collision between the galaxy and the intra-cluster medium (ICM)
of the Virgo cluster. In either case, such a large gas flow outside
the galaxy disk must be closely related to the evolution of NGC~4388
and the surrounding ICM, and thus detailed investigation of the VEELR
should provide important clues regarding the evolution of
galaxies and the ICM in clusters of galaxies.

To determine the nature and origin of the VEELR in NGC~4388 in
detail, we have carried out deep optical spectroscopic observations of the
VEELR filaments using the Subaru Telescope. We adopted a distance of
16.7 Mpc to NGC~4388 (Yasuda et al. 1997) throughout this study.

\section{OBSERVATIONS}

The observations were made with the FOCAS (Faint Object Camera And
Spectrograph; Kashikawa et al. 2002) attached at the Cassegrain focus
of the Subaru Telescope (Kaifu et al. 2000) on 2002 March 11.
We used the
multi-slit spectroscopy mode of FOCAS to obtain the spectra
of as many of the bright filaments of the VEELR as possible, given
their complicated spatial distribution. We used two slit masks, ``MOS mask 1'' and ``MOS
mask 2,'' and obtained the spectra of 40 filaments or clouds within
the 6{\arcmin} field of view of FOCAS. The PA of the slits of MOS mask
1 was 145$^{\circ}$, and that of MOS mask 2 was $-55^{\circ}$. The
layouts of the slits are shown in Figures 1 and 2.

The width of each slit was 0\farcs8 on the sky. We used a grating of
300 grooves/mm: the combination of the grating and the slit width gave
a resultant spectral resolving power of $\approx 650$ at 6600 {\AA}.
The detector is a mosaic CCD consisting of two 2K$\times$4K
MIT/Lincoln Lab. CCDs. The pixel scale of the CCDs is 0\farcs1 /
pixel. We binned 3 pixels in the spatial direction and 4 pixels in the
wavelength direction on the chip. Each spectrum covers the wavelength
range from 4700 {\AA} to 7500 {\AA}. We obtained six 1800-s exposures
and three 1800-s exposures for MOS mask 1 and MOS mask 2,
respectively, to improve the signal-to-noise ratio and to allow us to
avoid cosmic ray events and bad pixels of the detector by taking
median values in combining the spectra. The total exposure times were
therefore 3 hours and 1.5 hours for MOS mask 1 and MOS mask 2,
respectively.

Unfortunately, one CCD chip of the mosaic CCD has several defects and
bad columns and has a problem in charge transfer around the defects,
so that some regions of the CCD are covered by strong spurious patterns.
Although the total area of these spurious patterns is rather small as
a fraction of the total CCD area --- less than 1~\% ---, some spectra
were corrupted by this effect. We note all spectra that suffer from
this problem in Table 1 below.

We observed a spectrophotometric standard star, BD+33 2642, for flux
calibration. In observing the standard star we used a slit 2{\arcsec} wide.
The night was almost completely clear, but some patches of thin clouds
occasionally passed in front of the field of view. Seeing during the observing night was stable and
around 0\farcs7.

\section{DATA REDUCTION}

Data reduction was carried out using a special data reduction package
developed within IRAF\footnote{IRAF is distributed by the National
Optical Astronomy Observatories, which are operated by the Association
of Universities for Research in Astronomy, Inc., under cooperative
agreement with the National Science Foundation.} and IDL for FOCAS
multi-slit data (Yoshida et al. 2000; Kashikawa et al. 2002). After
the bias levels, which were estimated using the over-scan regions of
the CCD chips, were subtracted from each frame, flat fielding was
performed using the dome flat data. The field distortion of FOCAS was
corrected simultaneously with the flat-fielding.

We used the night-sky emission lines as wavelength calibration data
for $\lambda > 5700$ {\AA}, and Thorium arc line data at shorter
wavelengths. The wavelength calibration was accurate enough to
investigate the kinematics of the VEELR gas near the H$\alpha$ line;
the r.m.s.\ calibration error for $\lambda > 5500$ {\AA} is $\approx
0.2$ {\AA}, corresponding to 10 km s$^{-1}$ in radial velocity at
$\lambda = 6000$ {\AA}. The accuracy of absolute wavelength
calibration for the shortward wavelengths, i.e., $\lambda < 5500
${\AA}, was worse than for the red region; the calibration error is
about 1 {\AA} r.m.s., because the arc line data were obtained several
times during the observation and the positions of the spectra on the
CCDs are slightly shifted due to flexure of the FOCAS instrument
between the object exposure and the arc line exposure. Hence, we used
only H$\alpha$ data to measure radial velocities.

After subtracting sky emission from the data, we carried out flux
calibration using the standard star data and the standard extinction
curve at Mauna Kea. The flux calibration error, which was estimated
from the variation of the resultant fluxes of the calibrated spectra
of different frames, was sometimes as high as 30~\%. Most of this
inaccuracy was caused by the unstable sky conditions of the night of
the observations, as mentioned in the previous section. The flux
calibrated spectra were rescaled with respect to the frames showing
the highest count rate, then combined by taking the median value of
the frames. We summed the two-dimensional spectrum of each slit along
the direction of the slit length and made a one-dimensional spectrum.
The integration lengths along the slits are listed in column 3 of Table
1 in units of parsec. In some slits, there are at least two spatially
separated gas components, which differ from each other both in
kinematics or and in excitation. We extracted such components
separately from the two-dimensional spectrum if they were present.
Such components are indicated by adding the suffixes ``-1,'' ``-2,'' etc.\
to the slit IDs in Table 1. For example, the spectrum of the slit
M1-17 has two components with relative velocities of $-145$ km
s$^{-1}$ and $-438$ km s$^{-1}$, respectively: these are denoted
``M1-17-1'' and ``M1-17-2,'' respectively.

The emission-line characteristics (FWHMs, central wavelengths, and
intensities) were measured by fitting Gaussian functions. We fitted
the H$\alpha$ and [\ion{N}{2}] $\lambda\lambda$6548,6584 doublet
simultaneously, assuming the same recession velocity and velocity
dispersion for the three emission-lines and a fixed emission-line
intensity ratio of 3.0 for [\ion{N}{2}] $\lambda$6584/[\ion{N}{2}]
$\lambda$6548. We fitted the [\ion{S}{2}] $\lambda\lambda$6717,6731
doublet assuming the same recession velocity and velocity dispersion
for the two lines. We made the same assumption with a fixed line
intensity ratio of 3.0, for [\ion{O}{3}] $\lambda$5007/[\ion{O}{3}]
$\lambda$4959 in fitting H$\beta$ and
[\ion{O}{3}]$\lambda\lambda$4959,5007 doublet. Unfortunately, the
[\ion{O}{1}] $\lambda$6300 emission-line of the object falls near a
bright sky emission-line of [\ion{O}{1}] $\lambda$6364. The intrinsic
weakness of the [\ion{O}{1}] line, coupled with this problem,
prevented us from measuring the [\ion{O}{1}] emission-line parameters
accurately, except in rather bright cases.

\section{RESULTS}

The physical parameters of the VEELR filaments are tabulated in Table
1. The recession velocities of the filaments shown in column 5 are
those of H$\alpha$. As the spectral resolution of our spectra, $\sim
650$, is too low to allow accurate measurement of the velocity dispersion of the
emission-lines, we list the values for clearly resolved,
high signal-to-noise data in column 6. Instrumental broadening of
the emission line was roughly corrected using the following simple equation:
$\sigma_{true} = (\sigma_{obs}^2 - \sigma_{ins}^2)^{1/2}$, where
$\sigma_{obs}$ is the observed emission-line width and $\sigma_{ins}$ is
the instrumental line width. Column 7 of Table 1 shows the names of
individual clouds as given by YOS02.

The electron densities $N_{\rm e}$ were obtained from the intensity
ratios of [\ion{S}{2}] $\lambda$6717/[\ion{S}{2}] $\lambda$6731 under
the assumption that the ionized gas temperature is 10$^4$ K
(Osterbrock 1989). We found that the [\ion{S}{2}] $\lambda$6731 lines
of most filaments of the VEELR are too weak, relative to the
[\ion{S}{2}] $\lambda$6717 line, to allow accurate determination of $N_{\rm e}$:
that is, the $N_{\rm e}$s of the VEELR filaments are lower than $\sim
50$ cm$^{-3}$.

The emission-line intensity ratios of [\ion{O}{3}]
$\lambda$5007/H$\beta$, [\ion{N}{2}] $\lambda$6584/H$\alpha$,
[\ion{S}{2}] $\lambda$6717+$\lambda$6731/H$\alpha$ and [\ion{O}{1}]
$\lambda$6300/H$\alpha$ are summarized in Table 2.

\subsection{Velocity field}

The velocity field of the VEELR is shown in Figure 3. The velocities
of almost all the filaments measured are blue-shifted relative to the
systemic velocity of the galaxy. The velocities show
a very wide range ($\sim 700$ km s$^{-1}$) from $\sim -50$ km
s$^{-1}$ to over $-700$ km s$^{-1}$.

The overall velocity field of the VEELR is quite complicated and could
be dominated by significant turbulent motion. There is no smooth
velocity gradient across the region. Although we measured only a
fraction of the filaments in the region, several kinematic groups seem
to be represented (see Figure 4).

First, a string of relatively low velocity, $\sim -50$ km s$^{-1}$
-- $-200$ km s$^{-1}$, filaments (hereafter referred to as the ``LV
filaments'') was found along a line originating at around PA
65$^\circ$ and $r \approx 175${\arcsec} (14.2 kpc; M2-10) northeast
from the nucleus and extended along PA 55$^\circ$.

Second, a group of high velocity, $\sim -200$ km s$^{-1}$ -- $-450$
km s$^{-1}$, filaments (hereafter referred to as the ``HV filaments'')
are distributed over the region. The HV filaments can be broken down
into the following three groups: the ``NE-HV filaments,'' the ``N-HV
filaments,'' and the ``W-HV filaments.'' The NE-HV filaments are a bow-like string of filaments that extend from the cloud at PA
65$^\circ$, $r \approx 290${\arcsec} (around M2-14) to the north along
PA 10$^\circ$. The N-HV filaments are a group of clouds at PA
40$^\circ$, $r \approx 110${\arcsec} (around M1-5 and M2-6), PA
32$^\circ$, $r \approx 170${\arcsec} (M1-9 and M1-10) and PA
45$^\circ$, $r \approx 195${\arcsec} (M1-11 and M2-11). The W-HV
filaments consist of a bright, dumbbell-like cloud,
corresponding to the ``C26'' of YOS02, located at PA 95$^\circ$ and $r
\approx 125${\arcsec} (M1-1, M1-2 and M2-5), and a bright cloud at PA
70$^\circ$ and $r \approx 155${\arcsec} (M2-9), corresponding to 
``C25'' of YOS02.

Third, there is a group of very high velocity clouds/filaments
(hereafter referred to as the ``VHV clouds'') around the region where
PA$= 40^\circ$, $r \approx 145${\arcsec} (M1-7, M1-8, M2-7 and M2-8).
The typical radial velocity of the VHV clouds is $\sim -550$ km
s$^{-1}$, and one of these clouds even has a velocity of $\sim -750$
km s$^{-1}$. The VHV clouds are embedded in a string of filaments
that extend from PA 50$^\circ$, $r \approx 70${\arcsec} toward the
north-northeast (PA 15$^\circ$). This string of filaments contains the
N-HV filaments and the VHV clouds. The morphology of this string
suggests that the VHV clouds are part of one stream. If this were the
case, the high velocity motion of the VHV clouds could be interpreted
as a projection effect of the helical motion of a outflow stream.
However, the helical motion velocity of the stream would then reach up to
200 km s$^{-1}$, which is too fast to maintain the shape of the stream
unless there is a very strong confinement force, such as a strong
magnetic field. Thus, the greater fraction of their high velocities
cannot be interpreted as a projection of a helical motion of the
stream. Instead, the VHV clouds must be accelerated radially by some as yet undetermined
mechanism.

These names for the groups (the ``LV filaments,'' ``NE-HV filaments,''
``N-HV filaments,'' ``W-HV filaments,'' and ``VHV clouds'') are given
in column 8 of Table 1. In addition, the spectra of the NE
plume\footnote{The ``NE plume'' is the bright extra-planar
emission-line region extending up to $\sim 5$ kpc to the northeast of
the galaxy (e.g., Pogge 1988; Colina 1992; Petitjean and Durret 1993;
Veilleux et al. 1999a). The configuration of the extra-planar
emission-line region of NGC~4388 was illustrated in Figure 2 of
YOS02.} and the galaxy disk are labeled as ``NEp'' and ``disk'' in
column 8 of Table 1.

The positions and possible extensions of these filament/cloud groups
are overlaid on the velocity field of the VEELR in Figure 4. Figure 5
shows a position -- velocity diagram for the VEELR. The horizontal
axis of this diagram represents the distance from the nucleus in units
of kpc. Data points for different filaments/clouds groups are shown as
different symbols in this diagram (Figure 5). The groups of filaments
mentioned above can be seen to be well separated in position --
velocity space.

\subsection{Excitation of the filaments}

The emission-line spectra of the VEELR filaments are characterized by
forbidden line enhancement and are very different from those of
extragalactic \ion{H}{2} regions. Figure 6 shows some examples of the
spectra and Figures 7--9 show diagnostic diagrams for optical
emission-lines (Veilleux \& Osterbrock 1987). Two different ionization
model loci are over-plotted in Figures 7--9. The dotted lines
represent a power-law photoionization model calculated with the
photoionization code CLOUDY (CLOUDY 94.00: Ferland 1996). We used a
simple power-law continuum with a spectral index $\alpha$ of $-1.4$ as
the incident light, assuming a plane parallel cloud geometry with an
electron density $N_{\rm e}$ of 30 cm$^{-3}$ and solar metal
abundance. The dotted-dashed lines are the loci of the shock ionization
model of Dopita and Sutherland (1995). The loci shown here are for a
shock only model with no pre-existing magnetic field.

Figures 7--9 show that the emission-line ratios of almost all the
filaments are consistent with a power-law photoionization model with
solar metal abundance, suggesting that the filaments have
approximately solar abundance and are ionized by the power-law
ionizing continuum emerging from the nucleus of NGC 4388. This result
is consistent with the conclusions regarding the ionization mechanism of the VEELR
reported in previous studies (YOS02; Ciroi et al. 2003). The
ionization parameter $U$, which is the ratio of the number density of
ionizing photons to that of electrons, of a typical filament of the
VEELR was estimated to be of the order of $10^{-3}$ from the diagrams (Figure
7--9). The filaments showing relatively high $U$ ($\sim 10^{-2.5}$--$10^{-3}$)
tend to be distributed near the nucleus (i.e NE-plume and inner N-HV filaments),
whereas low $U$ ($\sim 10^{-3}$--$10^{-3.5}$) filaments are located far
from the nucleus (outer N-HV filaments and VHV filaments). This also
supports the AGN photoionization hypothesis.

Some filaments show low excitation spectra. Low excitation
emission-lines, such as [\ion{O}{1}] and [\ion{S}{2}], are enhanced in
the spectra of the W-HV filaments (``C26''). On the other hand,
[\ion{O}{3}] is relatively weak ([\ion{O}{3}]/H$\beta$ $\sim$ 1) in
these filaments, in contrast to other high excitation filaments. The
emission-lines of the W-HV filaments are also considerably broad (see
the bottom right panel of Figure 6). The data points for these regions
are distributed around the model locus of the shock model of Dopita
and Sutherland (1995) with a shock velocity of 200 -- 300 km s$^{-1}$.
These findings strongly suggest that the W-HV filaments are excited by
shocks. The W-HV filaments exist outside the ionization cone of the
NGC~4388 nucleus (Falcke, Wilson, and Simpson 1998; Veilleux et al.
1999a, YOS02). YOS02 remarked that the excitation mechanism of these
``out of the cone'' filaments is a mystery, pointing out that there is
no counterpart to C26 in the broad-band images (V and R$_C$ images),
so that it is not plausible that this filament is a bright \ion{H}{2}
region at the outskirts of the galaxy disk.

Shock may also contribute to the excitation of the other filaments.
Figure 10 shows the [\ion{N}{2}]/H$\alpha$ ratios of the VEELR
filaments as a function of velocity. This figure clearly shows that
the [\ion{N}{2}] line is enhanced in the VHV clouds (see also the
upper right panel of Figure 6). This velocity -- excitation coupling
indicates that shocks, induced by the rapid motion, can also play a
role in excitation of the filaments.

Ciroi et al. (2003) carried out a detailed analysis of the ionization of the
VEELR and the SW cone (Falcke, Wilson \& Simpson 1998; YOS02) using a
composite model of photoionization and shock heating. They used the
photometric data of YOS02 for analysis of the VEELR. The
ionization parameters they obtained for the VEELR filaments were
distributed in the range 10$^{-3}$ -- 10$^{-5}$. These values are
consistent with our results, although they are slightly lower.
This subtle discrepancy may be attributed to the contribution of shock
heating to ionization of the filaments. Ciroi et al. (2003) suggested
that shock heating gives rise to the ionization of the filaments, and
derived shock velocities in the range $\sim 30$ -- 150 km s$^{-1}$.
Note that these values are almost consistent with the shock velocities
suggested in the present study for the low excitation clouds of the VEELR.

\section{DISCUSSION}

\subsection{The origin of the VEELR gas}

YOS02 discussed the origin of the gas of the VEELR from a
morphological point of view, and concluded that there are two
possibilities for the origin of the VEELR: it is either the tidal
debris of a past minor merger, or the disk gas of NGC 4388 stripped by
the ram pressure of the hot ICM. Here, we examine these two scenarios
using our new observational results, and discuss the origin of the
VEELR gas.

\subsubsection{Minor merger tidal debris hypothesis}

As suggested in YOS02, NGC~4388 might have experienced at least one minor
merger in the past. A minor merger affects the dynamics of the primary
galaxy and leaves some imprints on its morphology (Hernquist \& Mihos
1995; Mihos et al. 1995). Peculiar morphological characteristics of
NGC~4388, such as its boxy bulge and central bar, and the faint hump
and tail that extend outside the disk, could have formed as a result
of the dynamic disturbance induced by a minor merger (YOS02). If the
VEELR were the tidal debris of a merger of a gas-rich dwarf galaxy and
NGC~4388, the gas in the dwarf should have been stripped by the tidal force
of NGC~4388 and the stripped gas would have been left near the path
of the dwarf's fall. If this were the case, the velocity of the VEELR gas
relative to NGC 4388 should not exceed the infall velocity of the
merging dwarf. However, the measured velocities of the VEELR filaments
are $-300$ -- $-400$ km s$^{-1}$, and some filaments show much higher
radial velocities, up to $-700$ km s$^{-1}$. These velocities are well
beyond the escape velocity  of NGC~4388 ($\sim \sqrt{2} \times v_{\rm rot}
\approx 250$ km s$^{-1}$, where $v_{\rm rot}$ is taken to be 180 km
s$^{-1}$, following Veilleux et al. 1999b) and so are too high to
apply the tidal debris scenario.

In addition, the turbulent nature of the velocity field of the VEELR
is not consistent with the minor merger hypothesis. The tidal tails
found in and around merging galaxies have smooth velocity fields and
mild velocity gradients of the order of 1 -- 10 km s$^{-1}$ kpc$^{-1}$
in general (e.g., Smith 1994; Hibberd et al. 1994; Hibberd and Yun
1999). For example, the Magellanic Stream, the nearest example of
tidal debris, has a velocity gradient of $\sim 300$ km s$^{-1}$ over a
few tens of kpc (Mathewson, Cleary \& Murray 1974). The velocity field
of the Magellanic Stream is also smooth and its kpc-scale velocity
dispersion does not exceed $\sim 100$ km s$^{-1}$. On the other hand,
the VEELR is highly turbulent (see Figures 3, 4). The velocities of
the filaments range over 700 km s$^{-1}$ and abrupt velocity changes,
with magnitudes of up to $\sim 300$ km s$^{-1}$, are seen over the
region.

Further, the emission-line spectra of the VEELR filaments suggest that
the metallicity of the VEELR gas is almost at the solar value: there is no
evidence that the gas is metal-poor. This is inconsistent with the
hypothesis that the VEELR gas was dragged out of a merged gas-rich
dwarf galaxy, because a typical gas-rich dwarf galaxy has a metal
abundance of 1/10 solar (cf. V\'ilchez \& Iglesias-P\'aramo 2003).

These results seem to disfavor the minor merger hypothesis. As
mentioned above, NGC~4388 might have experienced a minor merger
in the past (cf. YOS02), which may have given rise to part of the VEELR. 
However, it is difficult to explain most of the spectroscopic
properties of the VEELR gas only with the minor merger hypothesis.
Therefore, we concluded that it is unlikely that the majority of the
VEELR gas is tidal debris from a past minor merger.

\subsubsection{Ram pressure stripping hypothesis}

The characteristics of the VEELR described above can be naturally
explained by the ram pressure stripping hypothesis. 

As noted in
YOS02, the morphology of the VEELR strongly resembles some snapshots from
the inclined collision models of ram pressure stripping simulations
conducted by Abadi et al. (1999), Quilis et al. (2000), VCBD, 
and SS01. The one-sided elongation of
the VEELR suggests that NGC 4388 has a large transverse velocity toward
the southwest. The H$\alpha$ gas of the inner disk of the galaxy is
abruptly truncated at a distance of 5 kpc from the nucleus on the western
side, while on the eastern side of the galaxy faint ionized gas extends to the outside of the stellar disk (Figure 1a of YOS02). The
\ion{H}{1} gas distribution is also asymmetric in the same manner as
the H$\alpha$ gas (Cayatte et al. 1990). This peculiar gas
distribution supports the hypothesis that the galaxy is moving in a
southwesterly direction with a large transverse velocity and the disk
gas is blown out by ram pressure induced by this galactic motion.

The velocity field of the VEELR can also be interpreted in the context
of the ram pressure stripping scenario. While the recession velocity
of NGC~4388 is approximately 1500 km s$^{-1}$ larger than the systemic
velocity of the Virgo cluster, the galaxy is thought to be located at
the vicinity of the cluster core. Thus, NGC~4388 is moving in the Virgo
cluster ICM with a line-of-sight velocity of $\approx 1500$ km
s$^{-1}$ relative to the ICM. Hence, the blue-shifted velocity field
of the VEELR can be explained naturally: the collision between the
galaxy and the hot ICM strips the disk gas and the gas is blown in the
direction opposite to the motion of the galaxy. Assuming that the direction of the
extension of ram pressure-stripped gas is almost the same as the
direction of the ICM wind --- in other words, the stripped gas is
blown along the wind direction --- and that the inclination angle of
the VEELR with respect to our line of sight is 45$^{\circ}$, the
transverse velocity is also $\sim 1500$ km s$^{-1}$ and the total
infalling velocity of NGC 4388 toward the Virgo cluster center reaches
$\sim 2300$ km s$^{-1}$. 

Turbulent nature of the velocity field of the VEELR can also be
explained by high speed collision between the galaxy and the hot ICM.
Strong ram pressure from the ICM should induce turbulent motion in 
the stripped gas stream.

As described above, there seem to be a number of groups of
kinematically related filaments in the VEELR: the LV
filaments, the HV filaments, and the VHV clouds. VCBD presented some 
results of a series of numerical simulations
based on inclined collision models of ram pressure stripping. Edge-on
snapshots of a model of VCBD with an inclination
angle of 45$^{\circ}$ and a colliding velocity of $\sim 1000$ km
s$^{-1}$ show that the disk gas of the primary galaxy forms two
streams as it is stripped. One of the streams begins at the disk edge at
the nearside of the ICM wind and the other at the far side of the
wind. The latter stream is removed from the disk more rapidly than the former. 
Thus, it is suggested that the LV filaments correspond to the
former stream, while the HV and VHV clouds are embedded in the latter.

In addition, ram pressure stripping scnenario has no problem in explaining
high metallicity of the VEELR gas.
It is natural for the extended gas to have the same metallicity,
approximately Solar value,
to the galaxy disk, if the gas is stripped from the galaxy by ram pressure.

Therefore, the new spectroscopic data presented here led us to the
conclusion that it is most likely that the VEELR gas is the disk gas of
NGC~4388 stripped by the ram pressure of the hot ICM. A spatial
coincidence between the VEELR and the soft X-ray gas around NGC~4388,
which was found recently by detailed analysis of {\it Chandra}
archival data (Iizuka, Kunieda \& Maeda 2003) also supports the ram
pressure stripping hypothesis. If this is the case, we have detected ram
pressure-stripped gas in the form of warm ($T \sim 10^4$ K) ionized gas
(the VEELR) very far away from a cluster galaxy. The ionization of the
stripped gas by a powerful AGN, together with the deep imaging capability
of an 8-m class telescope, have made it possible to detect this faint
structure.

Although there have been many observational studies of ram pressure
stripping phenomena (e.g., Cayatte et al. 1990; Gavazzi et al. 1995;
Phookun \& Mundy 1995; Ryder et al. 1997; Kenney \& Koopmann 1999;
Vollmer et al. 2000; Vollmer, Braine et al. 2001; Bureau \& Carignan 2002),
the present study is the first to have revealed the detailed
morphology and kinematics of the ram pressure-stripped gas over a few
tens of kpc\footnote{Gavazzi et al. (2001) detected very large ($\sim
75$ kpc) warm ionized gas around the two irregular galaxies in the
cluster of galaxies Abell 1367, and concluded that these are ram
pressure-stripped material. This material is twice as large as the
VEELR of NGC~4388, but its detailed morphology and the kinematics are
not known.}. In fact, the situation of NGC~4388 is rather special, in
that the majority of the stripped gas happens to pass into the
ionizing radiation cone of the Seyfert nucleus, and as a consequence a
part of the stripped gas is ionized and visible as optical
emission-line gas. The case of NGC~4388 provides a rare opportunity to
investigate the warm phase of ram pressure-stripped gas in detail.

We discuss the VEELR in the context of the ram pressure stripping
hypothesis in the following sections.

\subsection{The VEELR and intracluster star formation activity}

In this section, we briefly discuss the fate of the VEELR gas and
examine the possibility of star formation in the VEELR.

The discovery of the VEELR indicates that some part of the ram
pressure-stripped gas survives throughout the stripping process and
does not evaporate for a significantly long time ($\sim 10^8$
yr)\footnote{The size and the radial velocity of the VEELR are $\sim
35$ kpc and $\sim 300$ km s$^{-1}$, respectively. Assuming that the
inclination angle of the VEELR is 45$^\circ$, we estimated the age of
the VEELR as $\sim 10^8$ yr}. VCBD pointed out that the
disk gas of the galaxy would be stripped in the form of an ensemble of
relatively dense cloudlets, each of which would be dense enough to
prevent its evaporation in the hot ICM during ram pressure stripping.
Those cloudlets would survive in the hot ICM for a long time ($t > 10^8$
yr). According to VCBD, the stripped cloudlets would
cool radiatively and become denser with time, so that molecules would
be formed in their cores. 

SS01 also pointed out
that the radiative cooling of ram pressure-stripped gas could be
important in cases in which the stripped gas density is higher than a
critical density, which is determined by the strength of the ram
pressure. They estimated that the critical density would be of the
order of $\sim 1$ cm$^{-3}$ for the Virgo cluster (SS01).
Although our spectroscopic results did not provide a definite value for
the electron densities of the VEELR filaments (they are too low to be
derived from [\ion{S}{2}] emission line ratios) YOS02 estimated the
r.m.s.\ electron density of each filament to be $\sim 0.5$ cm$^{-3}$
from H$\alpha$ photometry. Hence, assuming a typical volume filling
factor of interstellar ionized gas, $f_{\rm v} \sim 10^{-3} - 10^{-4}$
(e.g., Robinson et al. 1994), for the VEELR filaments, we estimate the
local densities of the line emitting clouds to be $\sim 10$ cm$^{-3}$.
This is high enough to make radiative cooling dominant in the thermal
balance of the stripped gas. The filamentary structure of the VEELR
also suggests that radiative cooling and thermal instability cause the
stripped gas to condense.

In the VEELR, there are a number of filaments that could be ionized
by the radiative shock induced by a collision between the disk ISM and
the hot ICM. Although the main ionization source should be the nuclear
power-law radiation, shock waves accompanied with rapid motion could
be responsible for the ionization of some HV filaments and VHV clouds. In
fact, for some VHV clouds that are distant from the nucleus, low
ionization emission-lines, such as [\ion{N}{2}] or [\ion{S}{2}], tend to
be enhanced, suggesting that shock heating plays some role in their
ionization. Successive shocks would condense the filaments
significantly. Therefore, the physical state of the VEELR filaments
allows them to form molecules internally, and to eventually form
stars.

Recently, Gerhard et al. (2002) found an intergalactic compact
star-forming region near NGC~4388. The location of the region is about
20 kpc north of the disk of NGC~4388. Although it is far away from the
main stream of the VEELR, it is worth noting that its distance from
the galaxy is comparable to that of the extension of the VEELR. This
region might be a gas clump that has split off from the main stream of
the VEELR. If this is the case, this region is the first known example
of star formation within ram pressure-stripped gas.

Our data suggest that ram pressure-stripped gas can 
survive for a long time after stripping and it might be
dense enough to cool radiatively.
As mentioned above, stars might be formed in such cool gas.
It is, however, no better than a speculation.
In order to obtain some conclusive results on intracluster star formation
and its relationship with ram pressure-stripping phenomenon,
deep observations of neutral gas, \ion{H}{1} or CO molecule,
would be crucial.

\subsection{Comparison with radio observations: interaction of disk star
formation with the VEELR}

We compared the deep H$\alpha$ and [\ion{O}{3}] images of NGC~4388
taken by YOS02 with radio observations to investigate
how the active star formation of NGC~4388 affects the extra-planar plasma
evolution.

Irwin et al. (1999) conducted VLA 20 cm and 6 cm observations of
NGC~4388 as a part of a series of studies of extended radio plasma
around edge-on galaxies. They found very extended, faint radio
emission on both sides of the disk of the galaxy at 20 cm. The
morphology of this emission is peculiar: it has a ``cracked'' X-shape.
It extends in three directions from the disk (to the northwest,
southeast, and southwest) but not in the northeasterly direction,
which is the direction of extension of the VEELR. The radio images of
Irwin et al.\ are overlaid on the H$\alpha$ image from YOS02 in
Figure 11b. We found faint extra-planar spurs in the H$\alpha$ image
of NGC 4388 (Figure 11d). The directions of these spurs agree well
with the direction of the extended radio emission. On the northeastern side
of the galaxy, the bright NE plume and the VEELR prevent the
identification of such faint structures.

Irwin et al. (1999) also found, at 6 cm, a striking extension
originating on the eastern side of the disk (10{\arcsec} east of the
nucleus) and extending northwards. The size of this extension is
comparable to that of the NE-HV filaments, whereas it is located
between the complex of the NE-HV filaments and the VHV clouds and the
NE plume (see Figure 11c). In other words, the extra-planar radio
plasma and the optical emission-line gas seem to be anti-correlated
in spatial distribution. This raises the question of the significance of this spatial anti-correlation. 

The faint radio emission seen at 20 cm and the northeastern extension
detected at 6 cm may be parts of a large X-shaped, starburst-driven
superwind (Chevalier \& Clegg 1985; Heckman, Armus \& Miley, 1990;
Strickland \& Stevens 2000) from the disk of NGC~4388. To the
northeastern side of the outflow, the radio-emitting plasma
may be accelerated by the turbulent motion of
the ram pressure-stripped ionized gas through some particle acceleration
mechanisms such as Fermi acceleration or shock acceleration
(cf. Webb et al. 2003; references are therein). The
spatial anti-correlation between the emission-line gas and the
radio-emitting plasma suggests an interaction between the two
components. This effect might change the
energy distribution of the radio-emitting plasma toward high energy,
causing the spectral index of the northeast extension to become
harder than the other wind plasma. Thus, the counterpart of the
northeast extension may be lost in the low energy band (i.e., 20 cm).

The interaction between the VEELR gas and the starburst superwind may also
accelerate the VEELR gas. The VHV clouds are located just outside
of the northeast top of the 6-cm radio emission. This spatial
coincidence supports the above suggestion. In other words, parts of the NE-HV
filaments may be accelerated by the superwind and form the VHV clouds.
This interaction would induce shock-waves in the VHV clouds and the
shocks would excite the gas.
The shock-like excitation property of the VHV clouds which was mentioned
in section 4.2 is naturally explained by this scenario.

An extended X-shaped structure has also been found in soft X-ray
emission. The {\it ROSAT} HRI image of NGC~4388, reduced by Colbert et
al. (1998), shows an X-shaped structure. Although the scale of this
emission is considerably smaller than the radio emission, the position
angles of the legs of the ``X''s are in good agreement. The size and
the luminosity of the HRI emission is comparable to those of the HRI
emissions of the starburst galaxy NGC~2146 (Armus et al. 1995).
Recently, the extra-planar soft X-ray emission is found to be more 
extended toward the south
of the galaxy at very faint level by a deep X-ray image of NGC~4388 
taken by {\it Chandra} (Iizuka, Kunieda, \& Maeda 2003).
As mentioned in section 5.1.2., the largest component of this faint 
X-ray emission is extended toward the north-eastern direction, which
is attributed to the ram pressure stripping.
The southern extension of this faint X-ray emission is seen like
an extension of the central ``X'' shaped X-ray emission of {\it ROSAT}.
These X-ray characteristics
also suggest that a superwind blows from the NGC~4388 disk and interacts
with the VEELR or the hot ICM wind.

The disk radio emission of NGC~4388 should be a result of active
star-forming regions in the galaxy disk\footnote{Active star-formation
in the disk of NGC~4388 is indicated not only by bright H$\alpha$ emission
but also by many chimney-like structures,
whch would be formed by supernovae explosions in the disk (cf. Norman and
Ikeuchi 1989), seen in the faint level of the emission-line image (see Figure 5 of 
YOS02).}. SS01
pointed out that the disk gas of the primary galaxy is significantly
compressed in the early phase of ram pressure stripping. The mechanism
of the compression is as follows: initially the disk gas is moved slightly 
in the direction of the ICM wind, but then the offset gas experiences the
gravitational force of the dark-matter halo and tends to fall back to
the galactic plane. Its motion is now in the opposite direction to the
ICM wind, so that the ICM wind and the gravitational force compress
the disk gas from both sides of the disk. This is the situation for a
face-on collision, i.e., one in which the galaxy is exposed to the ICM
wind face-on. In the case of an inclined collision, the disk gas is
also compressed in the disk plane by ram pressure. VCBD
have also discussed gas compression of this kind, occurring
before prompt stripping. Further, a third process, the gas
``annealing'' process discussed by SS01, may be
efficient in compressing the inner disk. In this process, angular
momentum transport by spiral waves induced in the ram pressure
stripping event leads to simultaneous expansion of the outer disk and
compression of the inner disk. The models of SS01
suggest that by the time the outer disk has been stripped, the inner
disk is significantly compressed.

A combination of the above processes must lead to strong compression
of the disk gas of NGC~4388. VCBD estimated the compression factor
of the gas surface density to be as high as 1.5 for the case of an edge-on collision. 
Other studies have also suggested that
gas compression by a factor of 2 -- 3 should occur in the early phases
of ram pressure stripping (e.g., Fujita and Nagashima 1999; SS01).
This high compression of disk gas should cause active
star formation in the disk. Assuming a Schmidt
law\footnote{$\Sigma_{SFR} = A\times \Sigma_{gas}^N$, where
$\Sigma_{SFR}$ is the star formation rate per unit area and $\Sigma_{gas}$
is the gas surface density.} (Schmidt 1959) with a slope $N \sim 1.5$
(Kennicutt 1998), we find that the star formation rate (SFR) increases
by a factor of up to 3 -- 5. In fact, the SFR of the NGC~4388 disk,
estimated from the H$\alpha$ emission, is $\sim 4$ M$_\odot$
yr$^{-1}$, which is consistent with the above value if we assume that
the SFR was $\sim 1$ M$_\odot$ yr$^{-1}$ before collision with the
ICM.

\subsection{Global Structure and Evolution of the Emission-line Region of NGC~4388}

Finally, we discuss the global structure of the extra-planar
emission-line region around NGC~4388. The following emission-line
regions are found in and around NGC~4388: (1) disk \ion{H}{2} regions,
(2) the NE plume and the SW cone (see Figure 2 of YOS02), (3) faint
extra-planar spurs (Figure 11d), and (4) the VEELR. Here, we propose an
evolutionary scenario in which the above emission-line regions can be
related to each other.

In this scenario we assume that the prime mover of the evolution of
the interstellar medium (ISM) of NGC~4388 is the fast collision
between the galaxy and the hot ICM of the Virgo cluster. Initially,
NGC~4388 was captured by the Virgo cluster during the cluster's mass
assembly process and started to fall into the central region of the
cluster. Once the galaxy began to experience ram pressure from the cluster ICM,
the ram pressure (and the opposing gravitational force) compressed the
disk gas. This would have taken place $\sim 10^7$ yr before the galaxy reached
the central region of the cluster. This gas compression caused
a starburst in the disk. Successive bursts of supernovae and stellar
winds from clusters of young massive stars heated the ISM, causing it
to expand into intra-cluster space: i.e., a starburst superwind was formed.

At the same time, the outer disk gas of NGC~4388 began to be stripped
by ram pressure. When the galaxy had passed through the cluster center,
the ram pressure gradually started to decline and the stripped gas
rotated slowly in the rotation direction of the host galaxy as a
result of its angular momentum. NGC~4388 probably collided with the
ICM to the southwestern side of the disk. As the disk rotation and
the ram pressure stripping force had the same direction on the western
side of NGC~4388, the western disk gas was the first to be dragged out
of the galaxy.

At the eastern side of the disk, where the directions of the disk
rotation and the ram pressure force were opposite to each other, the
disk gas lost its angular momentum due to deceleration by the ram
pressure. This disk gas thus started to fall in toward the nucleus.
In fact, this gas infall might have excited the nuclear activity,
although the triggering mechanism of AGN is not known in detail and
this is just one of several possible hypotheses. Recently, minor
merger events have been proposed as a promising mechanism for
triggering AGN activity (Taniguchi 1999; Kendall, Magorrian \& Pringle
2003). As pointed out in YOS02, the disk of NGC~4388 bears the marks
of a past minor merger. Hence, a minor merger might have occurred and
triggered the AGN at about the same time that NGC~4388 passed near the
cluster center.

To the northeast of the galaxy, the stripped gas formed an elongated
filamentary structure as a result of the combination of its rotation
and the ram pressure forces it experienced. According to VCBD, 
the stripped gas forms highly elongated structures like
the VEELR $\sim 10^8$ yr after the time when the ram pressure force
reaches its maximum. Therefore, NGC~4388 passed through the cluster
central region $\sim 10^8$ yr ago and is now receding from the center
of the Virgo cluster.

The interaction between the stripped gas and the starburst superwind
accelerated the gas and changed the energy spectrum of the
radio-emitting plasma. The radio jet from the AGN of NGC~4388 was
possibly accompanied by a warm gas outflow, and Veilleux et al.
(1999a) suggested that this outflow might have formed the NE plume.
However, we do not believe that the AGN outflow was the primary factor
in forming the NE plume, because (1) there is a large discrepancy
between the PA of the radio jet and that of the NE plume, (2) the NE
plume has a highly disturbed morphology compared to the winds seen in
other AGN wind cases, and (3) the velocity of the NE plume is
comparable to that of the NE-HV filaments. In addition, the morphology
of the NE plume resembles that of the NE-HV filaments. Therefore, we
conclude that most of the structure of the NE plume was formed by ram
pressure stripping (cf. Corbin, Baldwin \& Wilson 1988; Petitjean \&
Durret 1993). The AGN outflow possibly contributed to the formation of
the NE plume, but if so its contribution must have been small.

In the above scenario, the primary cause of the various phenomena
occurring in NGC~4388 is the fast collision of the galaxy with the hot
ICM. 
We summarize the structure of the emission-line regions around NGC~4388
as a schematic drawing in Figure 12.
The various phenomena in NGC~4388 --- the AGN, the starburst, and
the ram pressure stripping --- will last for only a comparatively
short time. After another several tens of millions of years, these
activities will cease or decay significantly. As the velocities of
many VEELR filaments exceed the escape velocity of NGC~4388, most of
the gas in the VEELR will escape into intracluster space. On the other
hand, the gas that is presently in the galaxy disk will remain there,
because the ram pressure force acting on it will decay. 
Therefore, NGC~4388 will be observed as an `anemic' spiral galaxy with
no peculiar activity several tens of millions of years from now.

\section{Summary}

Using the Subaru telescope, we carried out deep optical
spectroscopic observations of the very extended emission-line region
(VEELR) found around the Seyfert 2 galaxy NGC 4388 in the Virgo
cluster. The H$\alpha$ recession velocities of most of the filaments
in the VEELR are highly blue-shifted relative to the systemic velocity
of the galaxy. The velocity field is complicated and the range of 
velocities is considerably broad ($\sim 700$ km s$^{-1}$). There are a
number of kinematically and morphologically distinct groups of
filaments. We classified the filaments into the
following groups: the low velocity (``LV'') filaments, the high
velocity (``HV'') filaments, and the very high velocity (``VHV'')
clouds. The HV filaments consist of the following spatially separated subgroups: the N-HV filaments, the NE-HV filaments, and the W-HV
filaments. The emission-line ratios of the VEELR filaments are well
explained by power-law photoionization models with solar abundances,
suggesting that the Seyfert nucleus of NGC 4388 is the dominant
ionization source of the VEELR and that the VEELR gas has moderate
metallicity. Shock heating also contributes to the ionization of some
filaments. In particular, the western part of the HV filaments (the
``W-HV filaments'') may be excited predominantly by shocks.

We conclude that the VEELR was formerly the disk gas of NGC 4388,
which has been stripped from the disk by the ram pressure caused by
the interaction between the hot intra-cluster medium (ICM) and the
galaxy. The velocity field and the morphology of the VEELR closely
resemble snapshots taken from some numerical simulations of this
situation. In the case of NGC 4388, the ram pressure-stripped gas,
which would ordinarily be seen as extended
\ion{H}{1} filaments, happens to be exposed and ionized by the AGN
light, and so can be seen as optical emission-line gas. This special
situation makes it possible to study in detail environmental effects,
especially ram pressure stripping by the hot ICM, on the evolution of
cluster galaxies and intracluster medium.

Finally, we point out that filamentary/clumpy structure of the VEELR
suggests that ram pressure-stripped gas may be cooled radiatively and
be condensed to dense, small clouds as being extended into the
inter-galactic space.
In other words, most of the debris of ram pressure-stripping could be found
as an ensumble of small \ion{H}{1} filaments, if the filaments are
not ionized by some energetic source like AGNs.
Thus high spatial resolution, deep \ion{H}{1} survey may be 
important to make statistics of the ram pressure-stripping phenomena
and its impact to the galaxy evolution.

\acknowledgments

  We are grateful to the staff of the Subaru telescope for their kind
help with the observations. We wish to thank Curt Struck for his useful 
comments and the referee for a constructive report.
In addition, M. Y. thanks the staff of
Okayama Astrophysical Observatory for their encouragement during the
course of this work. 
This study was carried out using the facilities at
the Astronomical Data Analysis Center, National Astronomical
Observatory of Japan. This research made use of NASA's
Astrophysics Data System Abstract Service. This work was financially
supported in part by the Japan Society for the Promotion of Science
(Grant-in-Aid for Scientific Research No. 14204018).


\clearpage

\begin{figure}
\plotone{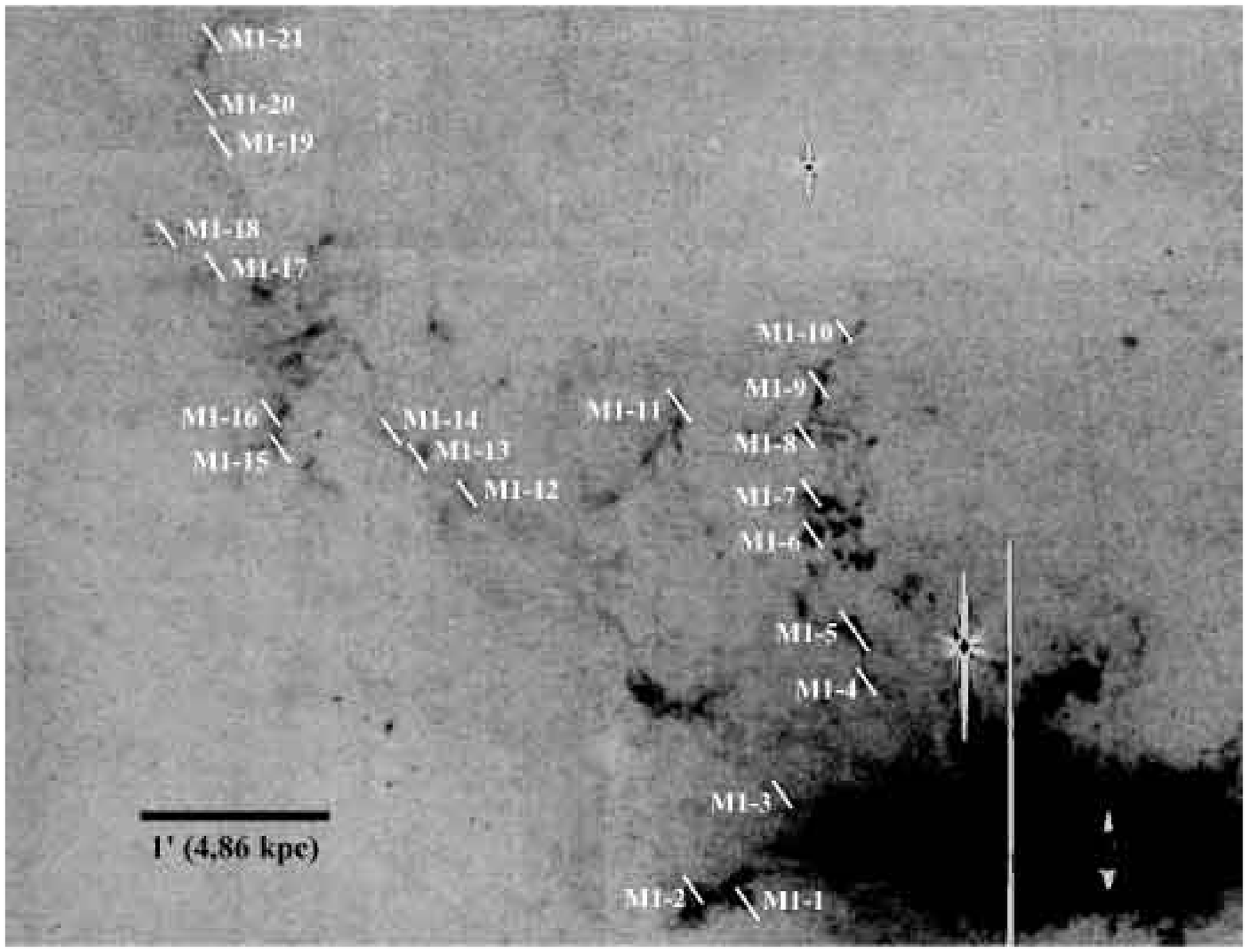}
\caption{
Figures 1 and 2 show the layouts of the slits.
This figure shows the slit layout of MOS mask 1.
The width of each slit is 0\farcs8 on the sky.
The scale bar indicates 1\arcmin.
\label{fig1}
}
\end{figure}

\begin{figure}
\plotone{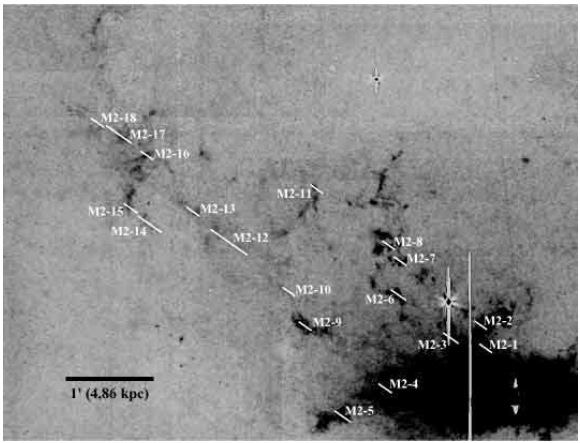}
\caption{
The slit layout of the MOS mask 2.
\label{fig2}
}
\end{figure}

\begin{figure}
\plotone{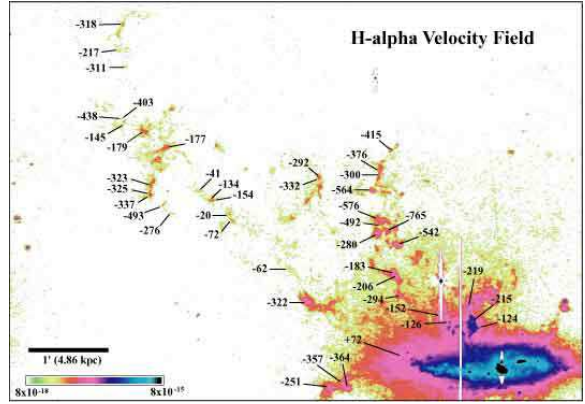}
\caption{
The velocity field of the VEELR of NGC~4388 overplotted on the H$\alpha$+[\ion{N}{2}]
image of YOS02.
The scale-bar at the left-bottom corner shows logarithmic scale of the surface
brightness of the H$\alpha$ emission in units of erg s$^{-1}$ cm$^{-2}$ arcsec$^{-2}$.
Radial velocities relative to the systemic velocity of the galaxy (2525 km s$^{-1}$)
are shown.
Note that almost all the filaments measured are blue-shifted relative to the galaxy.
The velocity field is complicated and there are significant velocity jumps in
some regions.
\label{fig3}
}
\end{figure}

\begin{figure}
\plotone{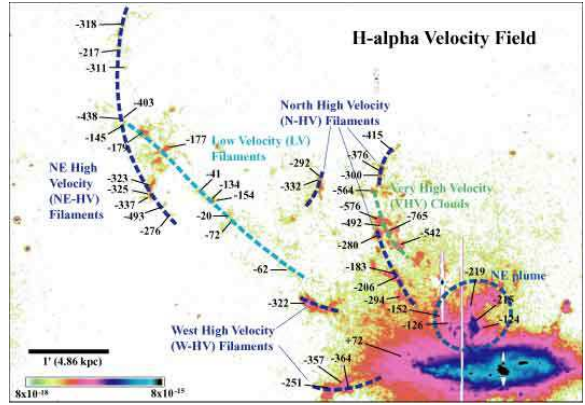}
\caption{
Schematic lines indicating individual groups of filaments/clouds suggested by
their morphological and kinematical characteristics (see the text) are overlaid on
the velocity field of the VEELR.
\label{fig4}
}
\end{figure}

\begin{figure}
\plotone{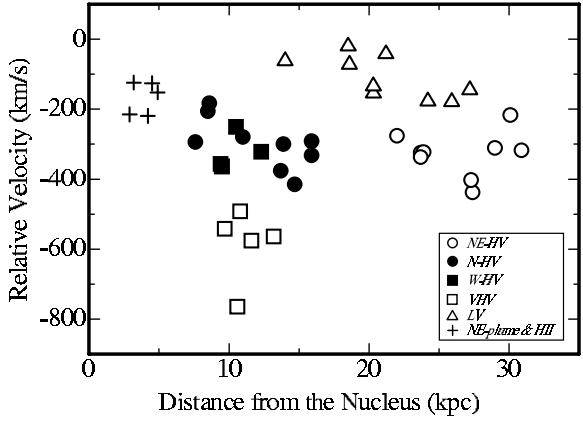}
\caption{
Distance -- velocity diagram for the VEELR. Different symbols indicate
different groups of filaments/clouds as follows: open circles, the
NE-HV filaments; filled circles, the N-HV filaments; 
filled squares, the W-HV filaments; open squares, the VHV clouds;
triangles, the LV filaments;
and crosses, the NE-plume and parts of the disk \ion{H}{2} regions.
\label{fig5}
}
\end{figure}

\begin{figure}
\plotone{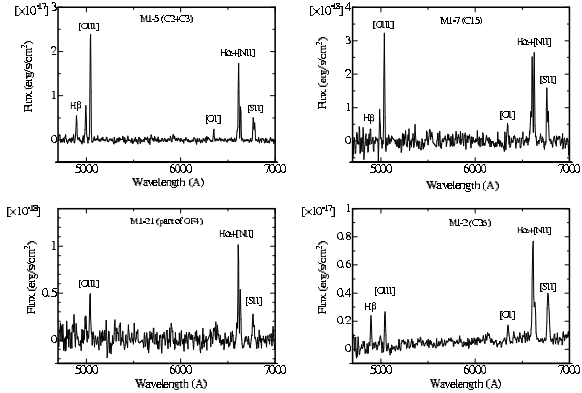}
\caption{
Some examples of spectra of filaments of the VEELR.
Top left: the spectrum of filaments C2 and C3 (YOS02) through slit M1-5.
Top right: the spectrum of filament C15 through slit M1-7.
Bottom left: the spectrum of the most distant filament measured in this study.
This spectrum was taken through slit M1-21.
Bottom right: the spectrum of filament C26 through slit M1-2.
\label{fig6}
}
\end{figure}

\begin{figure}
\plotone{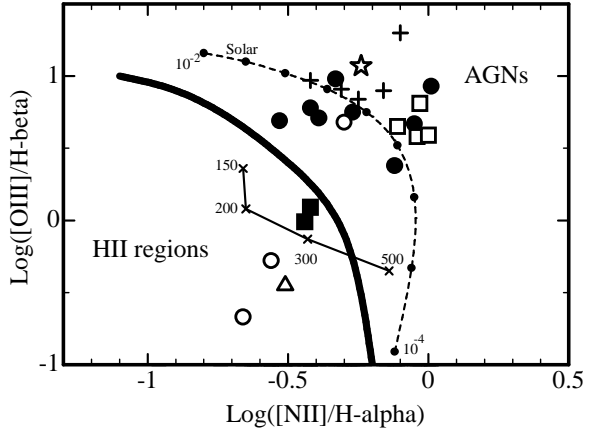}
\caption{
Fig.7 -- Fig.9 show emission-line diagnostic diagrams for the VEELR
filaments. Each axis of the diagrams has a logarithmic scale.
The symbols used in the diagrams are the same as in Figure 5. A star
symbol indicates data from the nuclear spectrum taken by Colina
(1992). This figure shows the [\ion{O}{3}]$\lambda$5007/H$\beta$ ratio
vs. the [\ion{N}{2}] $\lambda$6584/H$\alpha$ ratio. The thick solid
lines in the diagrams indicate the boundary between AGN-like spectra
and \ion{H}{2} region-like spectra.
The dashed lines are loci of a power-law photoionization model
calculated using CLOUDY (Ferland 1996). The electron density and the
spectral index $\alpha$ of the incident continuum of the model are 30
cm$^{-3}$ and $-1.4$, respectively. The metal abundance of the model is
the solar value. The filled circles on the loci represent a series of
ionization parameters $U$, from $U=10^{-4}$ to $U=10^{-2}$, with intervals
of $\log U=0.25$. The thin-solid lines are loci of the radiative
shock model (a ``shock only'' model with no magnetic field) of Dopita
and Sutherland (1995). The crosses on the loci indicate the shock
velocity, ranging from 150 km s$^{-1}$ to 500 km s$^{-1}$.
\label{fig7}
}
\end{figure}

\begin{figure}
\plotone{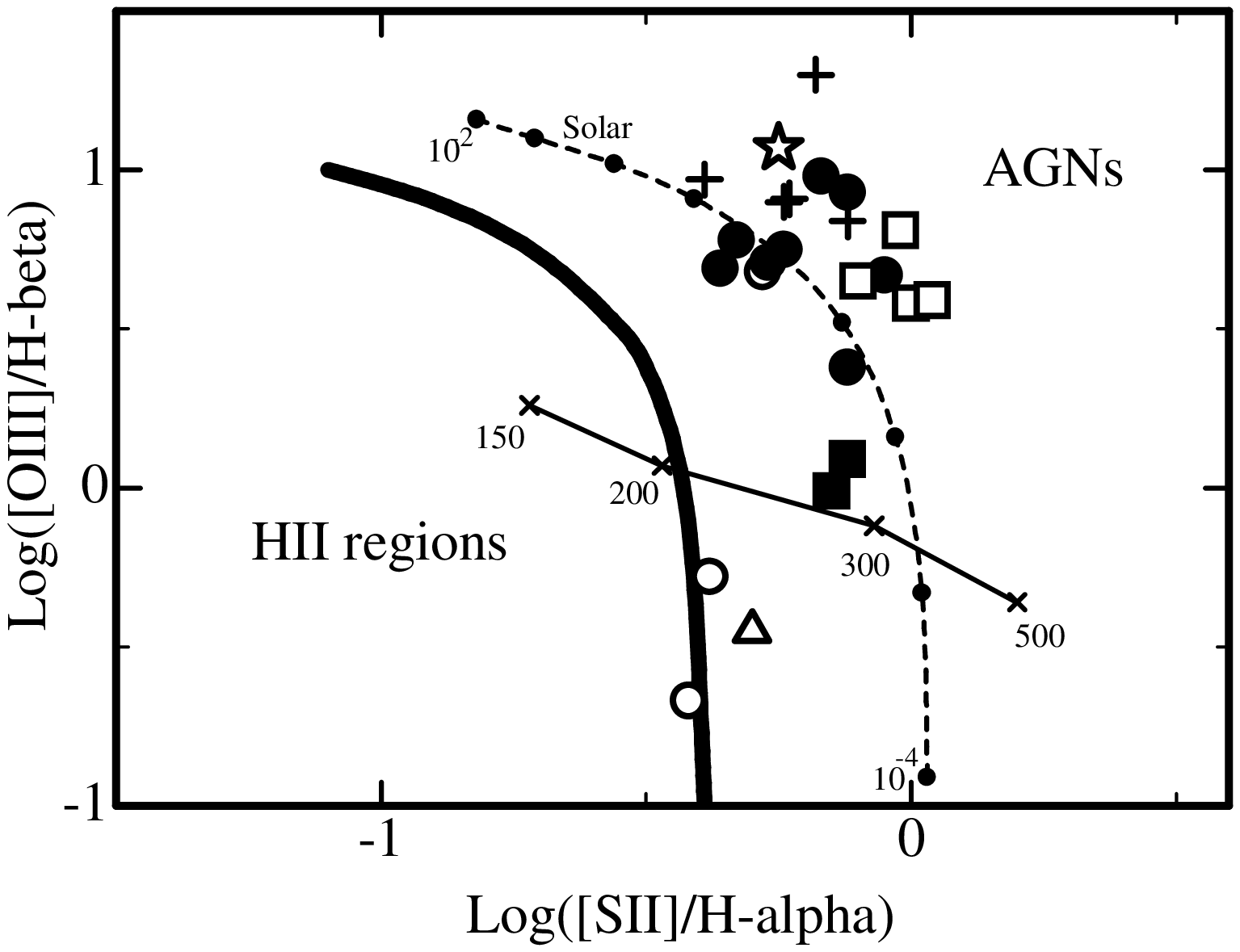}
\caption{
[\ion{O}{3}] $\lambda$5007/H$\beta$ ratio vs.
[\ion{S}{2}] $\lambda$6716+$\lambda$6731/H$\alpha$
ratio diagram.
The meanings of the symbols and lines are the same as in Figure 7.
\label{fig8}
}
\end{figure}

\begin{figure}
\plotone{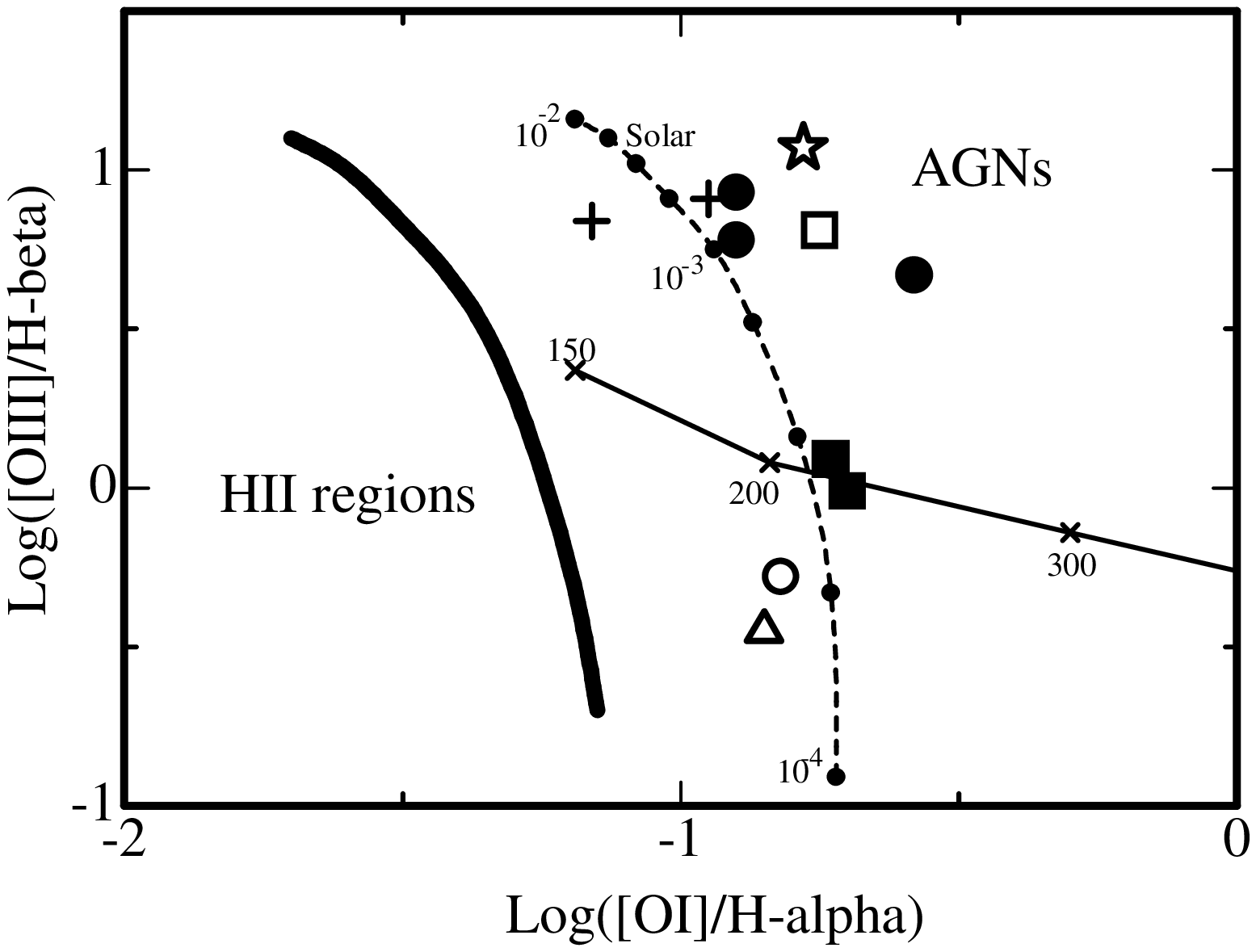}
\caption{
\label{Fig. 9}
[\ion{O}{3}] $\lambda$5007/H$\beta$ ratio vs.
[\ion{O}{1}] $\lambda$6300/H$\alpha$
ratio diagram.
The meanings of the symbols and lines are the same as in Figure 7.
\label{fig9}
}
\end{figure}

\begin{figure}
\plotone{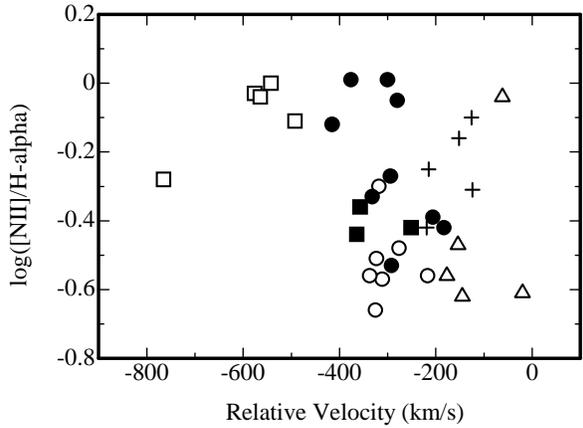}
\caption{
Velocity -- emission line intensity ratio (log([\ion{N}{2}]/[H$\alpha$])) diagram.
The meanings of the symbols are the same as in Figure 5.
\label{fig10}
}
\end{figure}

\begin{figure}
\plotone{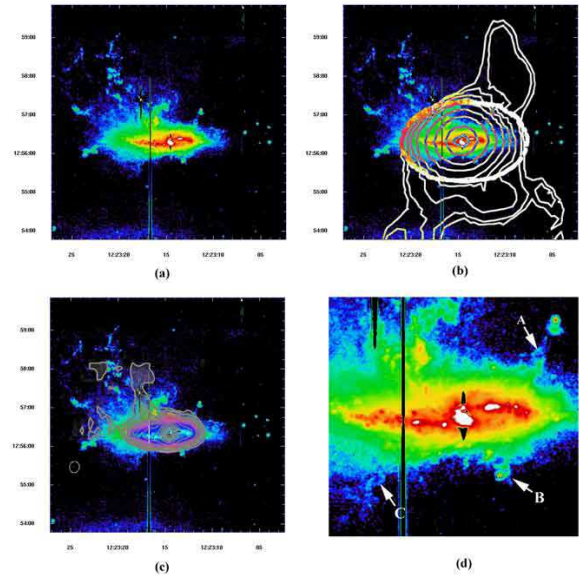}
\caption{
Comparison between the VEELR and the radio plasma distribution taken by Irwin et al. (1999).
a) The H$\alpha$ image (YOS02).
b) The 20 cm image overlaid on the H$\alpha$ image.
c) The 6 cm image overlaid on the H$\alpha$ image.
d) An expanded H$\alpha$ image of the central region of NGC~4388.
Three faint spurs (``A'', ``B'' and ``C'') are indicated by arrows.
\label{fig11}
}
\end{figure}

\begin{figure}
\plotone{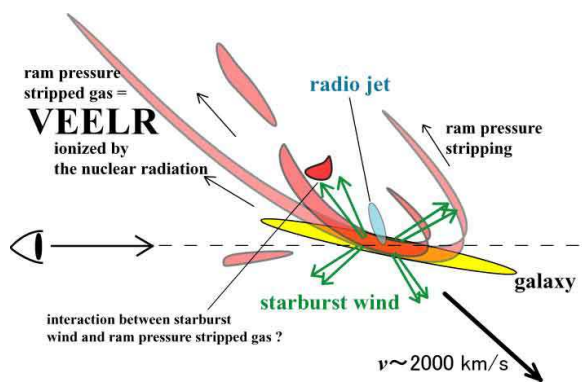}
\caption{
Schematic draw of the structure of the emission-line regions around NGC~4388.
\label{fig12}
}
\end{figure}

\clearpage

\begin{deluxetable}{lccrrrcc}
  \tabletypesize{\footnotesize}
  \tablewidth{0pt}
  \footnotesize
  \tablecaption{Physical Parameters of the VEELR Gas Clouds}
  \tablehead{
    \colhead{Slit ID} &
    \colhead{distance \tablenotemark{a}} &
    \colhead{length \tablenotemark{b}} &
    \colhead{$f_{\rm H\alpha}$ \tablenotemark{c}} &
    \colhead{$V_{\rm H\alpha}$ \tablenotemark{d}} &
    \colhead{$FWHM$ \tablenotemark{e}} &
    \colhead{Cloud ID \tablenotemark{f}} &
    \colhead{Group \tablenotemark{g}}
  }
  \startdata
    M1-1    & 9.5   & 150   & 14.6      & -364 $\pm19$  & 200       & C26 & W-HV \\
    M1-2    & 10.5 & 160    & 11.1      & -251 $\pm20$  & 530       & C26 & W-HV \\
    M1-4    & 7.6   & 230   & 3.3       & -294 $\pm22$  & 180       & C1 & N-HV \\
    M1-5    & 8.6   & 450   & 21.1      & -183 $\pm19$  & \nodata       & C2+C3 & N-HV \\
    M1-6    & 11.0 & 310    & 6.6       & -280 $\pm20$  & \nodata       & C11 & N-HV \\
    M1-7    & 11.6 & 190    & 3.6       & -576 $\pm21$  & \nodata       & C15 & VHV \\
    M1-8    & 13.2 & 260    & 4.6       & -564 $\pm20$  & \nodata       & C16 & VHV \\
    M1-9-1  & 13.7 & 50     & 0.3       & -376 $\pm87$  & \nodata       & C18 & N-HV \\
    M1-9-2  & 13.9 & 160    & 2.5       & -300 $\pm20$  & \nodata       & C18 & N-HV \\
    M1-10   & 14.7 & 150    & 2.8       & -415 $\pm21$  & \nodata       & C19 & N-HV \\
    M1-11   & 15.9 & 310    & 4.9       & -292 $\pm21$  & 80        & C22 & N-HV \\
    M1-12   & 18.5 & 110    & 0.7       & -20   $\pm32$ & 320       & part of OF1 & LV \\
    M1-13   & 20.3 & 160    & 1.6       & -154 $\pm22$  & \nodata       & part of OF1 & LV \\
    M1-14   & 21.2 & 100    & 0.4       & -41 $\pm47$   & \nodata       & part of OF1 & LV \\
    M1-15\tablenotemark{h}  & 23.7 & 160    & 4.1       & -325 $\pm22$  & 280       & part of OF2 & NE-HV \\
    M1-16\tablenotemark{h}  & 23.9 & 150    & 3.3       & -323 $\pm23$  & \nodata       & part of OF2 & NE-HV \\
    M1-17-1     & 27.2 & 80     & 0.6       & -145 $\pm26$  & \nodata       & part of OF2 & LV \\
    M1-17-2     & 27.4 & 100    & 0.8       & -438 $\pm71$  & \nodata       & part of OF2 & NE-HV \\
    M1-19\tablenotemark{h}  & 29.0 & 100    & 1.0       & -311 $\pm22$  & \nodata       & part of OF4 & NE-HV \\
    M1-20   & 30.1 & 80     & 0.3       & -217 $\pm30$  & \nodata       & part of OF4 & NE-HV \\
    M1-21   & 30.9 & 160    & 1.3       & -318 $\pm24$  & \nodata       & part of OF4 & NE-HV \\
            &   &   &      &        &  \\
    M2-1-1  & 2.9   & 150   & 21.0  & -215 $\pm19$  & \nodata       & part of NE plume & NEp \\
    M2-1-2  & 3.2   & 180   & 31.7  & -124 $\pm19$  & 30        & part of NE plume & NEp \\
    M2-2    & 4.2   & 440   & 15.8  & -219 $\pm20$  & 40        & part of NE plume & NEp \\
    M2-3-1  & 4.5   & 180   & 8.0       & -126 $\pm21$  & 140       & part of NE plume & NEp \\
    M2-3-2  & 4.9   & 80    & 2.5       & -152 $\pm21$  & \nodata       & part of NE plume & NEp \\
    M2-4    & 6.1   & 130   & 1.4       & +72 $\pm26$   & \nodata       & disk \ion{H}{2} region & disk \\
    M2-5    & 9.4   & 340   & 11.5  & -357 $\pm21$  & \nodata       & C26 & W-HV \\
    M2-6    & 8.5   & 320   & 14.7  & -206 $\pm20$  & 350       & C2+C3 & N-HV \\
    M2-7\tablenotemark{i}   & 9.7   & 190   & 4.9       & -542 $\pm20$  & \nodata       & C8 & VHV \\
    M2-8-1  & 10.6 & 100    & 1.3       & -765 $\pm31$  & \nodata       & C12 & VHV \\
    M2-8-2  & 10.8 & 100    & 2.5       & -492 $\pm23$  & 140       & C12 & VHV \\
    M2-9\tablenotemark{j}   & 12.3 & 60     & 0.9       & -322 $\pm132$     & 460       & C25 & W-HV \\
    M2-10\tablenotemark{h}  & 14.0 & 80     & 0.6       & -62   $\pm30$ & \nodata       & part of OF1 & LV \\
    M2-11   & 15.9 & 230    & 4.3       & -322 $\pm24$  & 230       & C22 & N-HV \\
    M2-12   & 18.6 & 230    & 2.0       & -72 $\pm31$       & 70        & part of OF1 & LV \\
    M2-13   & 20.3 & 80     & 0.7       & -134 $\pm25$  & \nodata       & part of OF1 & LV \\
    M2-14-1     & 22.0 & 130    & 1.1       & -276 $\pm33$  & 250       & part of OF2 & NE-HV \\
    M2-14-2     & 22.7 & 340    & 2.7       & -493 $\pm32$  & \nodata       & part of OF2 & NE-HV \\
    M2-15   & 23.7 & 160    & 4.4       & -337 $\pm20$  & 200       & part of OF2 & NE-HV \\
    M2-16   & 24.2 & 160    & 3.4       & -177 $\pm21$  & 290       & part of OF2 & LV \\
    M2-17   & 25.9 & 360    & 9.2       & -179 $\pm21$  & 280       & part of OF2 & LV \\
    M2-18   & 27.3 & 190    & 1.2       & -403 $\pm30$  & \nodata       & part of OF2 & NE-HV \\
  \enddata

 \tablenotetext{a}{distance from the nucleus in units of kpc.}
  \tablenotetext{b}{integration length along slit in units of pc.}
  \tablenotetext{c}{H$\alpha$ flux in units of 10$^{-17}$ erg s$^{-1}$ cm$^{-2}$.}
  \tablenotetext{d}{velocity relative to the systemic velocity of NGC~4388 (2525 km s$^{-1}$): km s$^{-1}$.}
  \tablenotetext{e}{full width at half maximum of the H$\alpha$ lines in units of km s$^{-1}$.}
  \tablenotetext{f}{The cloud identification of YOS02.}
  \tablenotetext{g}{The kinematical and morphological group identification
  defined in this work (see the text)}
  \tablenotetext{h}{CCD defect near [\ion{O}{3}] $\lambda$5007.}
  \tablenotetext{i}{CCD defect pattern at the edge of the slit.}
  \tablenotetext{j}{CCD defect pattern pollutes half of the spectrum along the slit.}

\end{deluxetable}

\begin{deluxetable}{lccrrrrc}
  \tabletypesize{\footnotesize}
  \tablewidth{0pt}
  \footnotesize
  \tablecaption{Emission-Line Intensity Ratios of the VEELR Gas Clouds}
  \tablehead{
    \colhead{Slit ID} &
    \colhead{Log([\ion{O}{3}]/H$\beta$)} &
    \colhead{Log([\ion{N}{2}]/H$\alpha$)} &
    \colhead{Log([\ion{S}{2}]/H$\alpha$)} &
    \colhead{Log([\ion{O}{1}]/H$\alpha$)}
  }
  \startdata
    M1-1    & -0.01 $\pm0.11$   & -0.44 $\pm0.06$   & -0.15 $\pm0.06$   & -0.70 $\pm0.10$   \\
    M1-2    & 0.09  $\pm0.10$   & -0.42 $\pm0.06$   & -0.12 $\pm0.06$   & -0.73 $\pm0.10$   \\
    M1-4    & 0.75 $\pm0.08$    & -0.27 $\pm0.06$   & -0.24 $\pm0.06$   & \nodata   \\
    M1-5    & 0.78 $\pm0.08$    & -0.42 $\pm0.06$   & -0.33 $\pm0.06$   & -0.90 $\pm0.10$   \\
    M1-6    & 0.67 $\pm0.08$    & -0.05 $\pm0.06$   & -0.05 $\pm0.06$   & -0.58 $\pm0.10$   \\
    M1-7    & 0.81 $\pm0.09$    & -0.03 $\pm0.06$   & -0.02 $\pm0.06$   & -0.75 $\pm0.11$   \\
    M1-8    & 0.58 $\pm0.08$    & -0.04 $\pm0.06$   & 0.00  $\pm0.06$   & \nodata   \\
    M1-9-1  & \nodata       & 0.01  $\pm0.07$   & -0.14 $\pm0.08$   & \nodata   \\
    M1-9-2  & 0.93 $\pm0.09$    & 0.01 $\pm0.06$    & -0.12 $\pm0.06$   & -0.90 $\pm0.11$   \\
    M1-10   & 0.38 $\pm0.09$    & -0.12 $\pm0.06$   & -0.12 $\pm0.06$   & \nodata   \\
    M1-11   & 0.69 $\pm0.09$    & -0.53 $\pm0.06$   & -0.36 $\pm0.06$   & \nodata   \\
    M1-12   & \nodata       & -0.61 $\pm0.08$   & \nodata   & \nodata   \\
    M1-13   & \nodata       & -0.47 $\pm0.06$   & -0.25 $\pm0.06$   & \nodata   \\
    M1-14   & \nodata       & \nodata   & \nodata   & \nodata   \\
    M1-15   & \nodata   & -0.66 $\pm0.06$   & -0.42 $\pm0.06$   & \nodata   \\
    M1-16   & \nodata   & -0.51 $\pm0.06$   & -0.26 $\pm0.06$   & \nodata   \\
    M1-17-1     & \nodata   & -0.62 $\pm0.07$   & -0.09 $\pm0.06$   & \nodata   \\
    M1-17-2     & \nodata   & \nodata   & -0.37 $\pm0.07$   & \nodata   \\
    M1-19   & \nodata       & -0.57 $\pm0.06$   & -0.30 $\pm0.06$   & \nodata   \\
    M1-20   & \nodata       & -0.56 $\pm0.08$   & \nodata   & \nodata   \\
    M1-21   & 0.68 $\pm0.11$    & -0.30 $\pm0.06$   & -0.28 $\pm0.06$   & \nodata   \\
            &       &       &       &           \\
    M2-1-1  & 0.84 $\pm0.08$    & -0.25 $\pm0.06$   & -0.12 $\pm0.06$   & -1.16 $\pm0.11$   \\
    M2-1-2  & 0.91 $\pm0.08$    & -0.31 $\pm0.06$   & -0.23 $\pm0.06$   & -0.95 $\pm0.10$   \\
    M2-2    & 0.97 $\pm0.08$    & -0.42 $\pm0.06$   & -0.39 $\pm0.06$   & \nodata   \\
    M2-3-1  & 1.30 $\pm0.09$    & -0.10 $\pm0.06$   & -0.18 $\pm0.06$   & \nodata   \\
    M2-3-2  & 0.90 $\pm0.08$    & -0.16 $\pm0.06$   & -0.24 $\pm0.06$   & \nodata   \\
    M2-4    & \nodata   & \nodata   & -0.55 $\pm0.08$   & \nodata   \\
    M2-5    & \nodata       & -0.36 $\pm0.06$   & -0.23 $\pm0.06$   & \nodata   \\
    M2-6    & 0.71 $\pm0.08$    & -0.39 $\pm0.06$   & -0.27 $\pm0.06$   & \nodata   \\
    M2-7    & 0.59 $\pm0.08$    & 0.00 $\pm0.06$    & 0.04  $\pm0.06$   & \nodata   \\
    M2-8-1  & \nodata       & -0.28 $\pm0.06$   & -0.14 $\pm0.06$   & \nodata   \\
    M2-8-2  & 0.65 $\pm0.09$    & -0.11 $\pm0.06$   & -0.10 $\pm0.06$   & \nodata   \\
    M2-9    & \nodata       & -0.15 $\pm0.07$   & \nodata   & \nodata   \\
    M2-10   & \nodata       & \nodata       & \nodata       & \nodata       \\
    M2-11   & 0.98 $\pm0.12$    & -0.33 $\pm0.06$   & -0.17 $\pm0.06$   & \nodata   \\
    M2-12   & \nodata   & -0.74 $\pm0.09$   & -0.25 $\pm0.06$   & \nodata   \\
    M2-13   & \nodata       & -0.96 $\pm0.10$   & -0.24 $\pm0.06$   & \nodata   \\
    M2-14-1     & \nodata       & -0.48 $\pm0.07$   & -0.37 $\pm0.07$   & \nodata   \\
    M2-14-2     & \nodata   & -0.63 $\pm0.08$   & -0.35 $\pm0.06$   & \nodata   \\
    M2-15   & -0.28 $\pm0.16$   & -0.56 $\pm0.06$   & -0.38 $\pm0.06$   & -0.82 $\pm0.11$   \\
    M2-16   & \nodata       & -0.56 $\pm0.06$   & -0.29 $\pm0.06$   & \nodata   \\
    M2-17   & -0.45 $\pm0.21$   & -0.51 $\pm0.06$   & -0.30 $\pm0.06$   & -0.85 $\pm0.11$   \\
    M2-18   & \nodata   & \nodata   & 0.02  $\pm0.06$   & \nodata   \\
  \enddata

\end{deluxetable}

\end{document}